# Microwave-free dynamic nuclear polarization via sudden thermal jumps

Carlos A. Meriles[1,2,†] and Pablo R. Zangara[3,4]

[1]*Department. of Physics, CUNY-City College of New York, New York, NY 10031, USA.*
[2]*CUNY-Graduate Center, New York, NY 10016, USA.*
[3]*Universidad Nacional de Córdoba, Facultad de Matemática, Astronomía, Física y Computación, Córdoba X5000HUA, Argentina.*
[4]*CONICET, Instituto de Física Enrique Gaviola (IFEG), Córdoba X5000HUA, Argentina.*

[†]*Corresponding author. E-mail: cmeriles@ccny.cuny.edu.*

Dynamic Nuclear Polarization (DNP) presently stands as the preferred strategy to enhance the sensitivity of nuclear magnetic resonance measurements, but its application relies on the use of high-frequency microwave to manipulate electron spins, an increasingly demanding task as the applied magnetic field grows. Here we investigate the dynamics of a system hosting a polarizing agent formed by two distinct paramagnetic centers near a level anti-crossing. We theoretically show that nuclear spins polarize efficiently under a cyclic protocol that combines alternating thermal jumps and radio-frequency pulses connecting hybrid states with opposite nuclear and electronic spin alignment. Central to this process is the difference between the spin-lattice relaxation times of either electron spin species, transiently driving the electronic spin bath out of equilibrium after each thermal jump. Without the need for microwave excitation, this route to enhanced nuclear polarization may prove convenient, particularly if the polarizing agent is designed to feature electronic level anti-crossings at high magnetic fields.

The ability to produce order from disorder is perhaps best captured through the notion of a heat engine, extracting mechanical work — a more organized form of energy — from alternately coupling to hot and cold reservoirs. In magnetic systems, Zeeman order is one such form of energy with low entropic content, and so it is natural to wonder whether a thermal cycle can be exploited to induce spin alignment. That this is actually the case is indirectly suggested by recent caloritronics experiments where spin polarization in a ferromagnet emerges from charge flow across a thermal gradient (the so-called spin Seebeck effect[1]). Yet, extending the governing principles to more general systems remains an outstanding problem, particularly if the polarization target is the nuclear spin ensemble of a non-ferromagnetic, non-conductive material host.

Admittedly, there is a long-standing, intimate connection between thermal equilibrium and nuclear spin order, already present in Overhauser's famous proposal to dynamic nuclear polarization[2] (DNP). Indeed, the key to this method lies in the ability to steer the electron spin reservoir away from thermal equilibrium through continuous microwave (mw) excitation; equally critical is the built-in asymmetry between the single and double quantum transition rates governing relaxation in the combined electron/nuclear spin system[3]. Although the search for alternative methods to actively polarize nuclear spins has grown to become itself an active field of research[4-8], microwave-based schemes — relying on the Overhauser effect or other mechanisms such as the solid effect, thermal mixing, or the cross effect — are today the most widespread[9]. In all these methods — including those where spin polarization is optically[10-12] or photo-chemically[13,14] pumped — the sample remains in contact with a thermal bath at a fixed temperature, hence suggesting there is room for other, conceptually different routes to DNP.

Here, we study a system comprising two dipolarly coupled paramagnetic centers featuring different spin-lattice relaxation times near a level anti-crossing. By implementing a thermal cycle featuring sudden temperature jumps, we show the spin of hyperfine-coupled nuclei can be dynamically polarized solely with the use of radio-frequency (rf) pulses adjusted to address a pre-selected pair of states with opposite nuclear (and electronic) spin orientations. Under steady state conditions, the limit nuclear polarization that emerges grows with the electron spin population change induced by the thermal jump.

To introduce some of the key ideas, we first consider the schematic in Fig. 1a comprising two separate containers $\sigma$, $\sigma'$, each conducting heat from a surrounding thermal bath with characteristic rates $W_\sigma < W_{\sigma'}$. We assume each container encloses two classes of particles, interconverting from one type to the other at a rate $W_\nu \ll W_\sigma, W_{\sigma'}$. Further, we ask that the equilibrium (combined) number of particles in each enclosure be proportional to the outside bath temperature $\mathcal{T}$, and assume we have at our disposal an external means ('gate') to quickly convert one type of particle into the other provided we also switch the containers they are in.

To "polarize" the containers — i.e., to increase the fractional number of one class of particle over the other — we implement the protocol in Fig. 1b: Starting from equilibrium at a lower temperature $\mathcal{T}$ — where both containers enclose the same (low) number of particles of either type, "stage 1" in the schematic — we quickly heat up the bath to a higher temperature $\mathcal{T}'$. If the time scale of the thermal jump $\Delta t_\uparrow$ is sufficiently short (i.e., when $\Delta t_\uparrow \lesssim W_{\sigma'}^{-1}$), a transient imbalance emerges between the particle populations in each container ("stage 2" in Fig. 1b), simply because only one container can thermalize with the bath. At this point, therefore, one can opt to generate polarization of one sign or the other by selectively activating a gate ("stage 3"). Note that the interconversion of particles within each container is slow, implying this polarization is preserved as the system cools back down ("stage 4" in Fig. 1b).

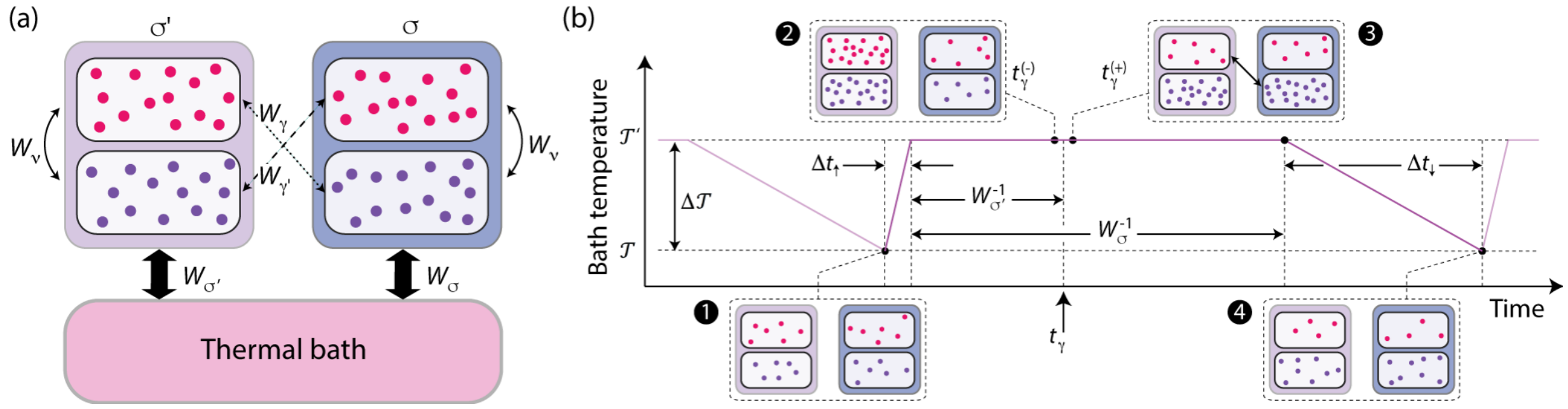

**FIG. 1: Dynamic polarization through thermal cycling.** (a) We consider two containers exchanging heat with a thermal bath at distinct rates $W_\sigma$, $W_{\sigma'}$. Each container encloses two classes of particles — red and purple — inter-converting at a rate $W_v \ll W_\sigma, W_{\sigma'}$; selective transformation of one class into the other can also be carried out externally at rates $W_\gamma \approx W_{\gamma'} \gg W_\sigma, W_{\sigma'}$, provided particles also switch containers (dotted and dashed arrows). In a given container, the total number of particles depends on the container temperature. (b) Starting from a configuration where the number of red and blue particles are the same (stage 1), a sudden thermal jump increases the number of particles in the left container first, thus allowing a gate pulse at $t_\gamma$ to increase the fractional content of blue particles (stages 2 and 3). Since $W_v$ is comparatively very slow, the imbalance remains when the bath returns to the original temperature (stage 4).

Figure 2a introduces a physical realization of the above model system in the form of a spin set comprising a hyperfine-coupled nuclear spin $I = 1/2$ and two paramagnetic centers, $S = 1$ and $S' = 1/2$, themselves interacting via a dipolar coupling $\mathcal{J}_d$. We write the system Hamiltonian as

$$H = \Delta S_z^2 - \gamma_e B S_z - \gamma_e B S_z' - \gamma_n B I_z + H_d(S, S') + A_{zz} S_z I_z + A_{zx} S_z I_x \,, \quad (1)$$

where $\Delta$ is the crystal field acting on spin $S$, and $A_{zz}$ ($A_{zx}$) denotes the secular (pseudosecular) hyperfine coupling constant on the nuclear spin, $\gamma_e$ ($\gamma_n$) is the electronic (nuclear) gyromagnetic ratio, and $H_d(S, S')$ is the dipolar coupling Hamiltonian between spins $S$ and $S'$[15]. For simplicity, we impose the same (scalar) gyromagnetic ratio $\gamma_e$ to both electron spin species, but assume different spin-lattice relaxation times $W_S^{-1} \equiv T_{1e}^{(S)} > T_{1e}^{(S')} \equiv W_{S'}^{-1}$. The latter is, in general, warranted because the presence of a crystal field in one of the paramagnetic centers creates an asymmetry

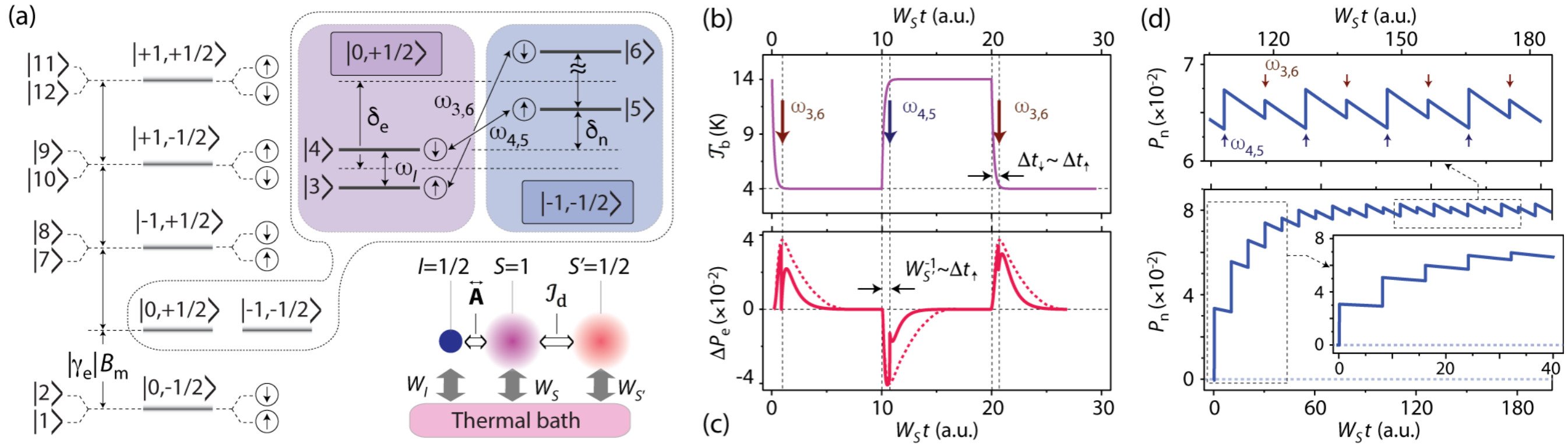

**FIG. 2: Thermal cycling near a level anti-crossing.** (a) Energy level diagram for a spin set formed by two paramagnetic centers with spin numbers $S = 1$, $S' = 1/2$, and a nuclear spin $I = 1/2$ in the vicinity of a level anti-crossing. Each spin couples to the thermal bath with characteristic spin-lattice relaxation times $T_{1e}^{(S)} \equiv W_S^{-1}$, $T_{1e}^{(S')} \equiv W_{S'}^{-1}$, and $T_{1n} \equiv W_I^{-1}$, respectively. (b) Bath temperature $\mathcal{T}_b$ as a function of the normalized time $W_S t$; arrows indicate the application of rf $\pi$-pulses at frequencies $\omega_{45}$ ($\omega_{36}$) during heating (cooling) intervals. (c) Electron spin population difference $\Delta P_e \equiv P_{56} - P_{34}$ as a function of the normalized time; red dotted lines indicate the response in the absence of rf pulses. (d) Nuclear spin polarization as a function of $W_S t$. The upper (lower) insert is a zoomed-out view of the response at late (early) times. Throughout these calculations, we assume $\Delta t_\uparrow = \Delta t_\downarrow = T_{1e}^{(S')} = T_{1e}^{(S)}/5 = T_{1n}/1000 = 10$ ms, $\mathcal{J}_d = 5$ MHz, $A_{zz} = A_{zx} = 10$ MHz, $\delta_e = 30$ MHz, $\delta_n = 10$ MHz, $\Delta = 100$ GHz, and $|\gamma_e| B_m = 50$ GHz; under the above conditions, $\omega_{4,5} = \delta_n = 10$ MHz, $\omega_{3,6} \simeq 58$ MHz..

between the spin-lattice relaxation channels at play for each spin species.

To recreate the polarization protocol in Fig. 1b, we bring the externally applied magnetic field $B$ to a value near the "matching condition", $|\gamma_e|B_m \equiv \Delta/2$, where the energy splitting between states $|-1\rangle$ and $|0\rangle$ in spin $S$ approximately coincides with the spacing between states $|\pm 1/2\rangle$ in $S'$. While there is some freedom in selecting the exact value of the magnetic field, a practical range emerges from a subtle interplay: On the one hand, we must make the energy detuning $\delta_e = |\gamma_e|(B - B_m)$ greater than a minimum value $\delta_e^{(\min)}$ so that no spontaneous electron-nuclear flip-flops between states $|4\rangle$ and $|5\rangle$ (or $|3\rangle$ and $|6\rangle$) can take place (see Fig. 2a for notation). Conversely, we must make $\delta_e$ smaller than an upper limit $\delta_e^{(\max)}$ so as to ensure reasonably high transition probabilities in the presence of resonant excitation at frequencies $\omega_{4,5}$ or $\omega_{3,6}$. Although nominally forbidden, these transitions — by construction, within the radio-frequency range — activate near the level anti-crossing as a result of state hybridization from inter-spin dipolar and hyperfine couplings, as demonstrated recently[16]. In passing, we caution that the use of exact $\pi$-pulses is not strictly mandatory as the required population exchange could be attained through longer, "saturation" pulses, or via rapid field sweeps that transiently align the energies of the relevant pair of states[17].

Figure 2b displays the results from numerical modeling in the case of a spin set featuring hyperfine and dipolar couplings of order 1–10 MHz, typical in organic systems; for presentation purposes, we impose a moderately large crystal field $\Delta = 100$ GHz (corresponding to an approximate matching magnetic field $B_m \approx 1.8$ T), and assume that both the heating and cooling times coincide with the (shorter) spin-lattice relaxation time of spin $S'$, i.e., $\Delta t_\uparrow = \Delta t_\downarrow = T_{1e}^{(S')}$; as discussed later, this condition appears very much compatible with the cryogenic conditions characteristic in current DNP protocols[18].

We readily map the dynamics introduced in Fig. 1 to the spin system at hand when we consider the population difference $\Delta P_e = P_{5,6} - P_{3,4}$ between the integrated electron spin populations $P_{2j-1,2j}$, $j = 2, 3$, as a function of the fractional time $W_S t$ (Fig. 2c). In thermal equilibrium, $\Delta P_e \approx 0$ given the small energy differences (caused by nuclear Zeeman and hyperfine interactions) at the assumed (nearly matching) field. This is no longer the case, however, after a sudden temperature jump because, unlike states $|5\rangle$ and $|6\rangle$, states $|3\rangle$ and $|4\rangle$ can quickly exchange population with ground states $|1\rangle$ and $|2\rangle$ through a flip of spin $S'$, thus inducing a transient population imbalance. For example, during a temperature increase, states $|3\rangle$ and $|4\rangle$ gain population from $|1\rangle$ and $|2\rangle$ whereas the occupancy of states $|5\rangle$ and $|6\rangle$ diminishes from exchange with states $|7\rangle$ and $|8\rangle$, hence leading to $\Delta P_e < 0$. Correspondingly, positive (negative) nuclear polarization follows from the application of an inversion rf pulse at $\omega_{4,5}$ ($\omega_{3,6}$). Further, provided the warm-up and cool-down times $\Delta t_\uparrow, \Delta t_\downarrow$ are equally fast — the case in Fig. 2b — nuclear polarization of the same sign can be produced following either jump upon switching the rf from one frequency to the other (Fig. 2d).

More rigorously, we can derive an approximate expression for the nuclear polarization $P_n = \sum_{j=1}^{6}\left(P_{2j-1} - P_{2j}\right)$ when we consider the limit case $\Delta t_\uparrow, \Delta t_\downarrow \lesssim T_{1e}^{(S')} \ll T_{1e}^{(S)}$. Assuming the system is initially in equilibrium at temperature $\mathcal{T}$, the transient integrated population $P_{3,4}^{(\mathrm{tr})}$ following a time $t_\gamma^{(-)} \gtrsim T_{1e}^{(S')}$ after a jump to temperature $\mathcal{T}'$ is given by

$$P_{3,4}^{(\mathrm{tr})} \approx \left(P_{1,2}^{(\mathrm{eq})}\Big|_{\mathcal{T}} + P_{3,4}^{(\mathrm{eq})}\Big|_{\mathcal{T}}\right)\frac{\exp(-\beta_e')}{(1+\exp(-\beta_e'))}, \quad (2)$$

where $\beta_e' \equiv |\gamma_e|B/k_B\mathcal{T}'$, and $k_B$ denotes Boltzmann's constant. Similarly, the integrated population in states $|5\rangle$ and $|6\rangle$ can be cast as

$$P_{5,6}^{(\mathrm{tr})} \approx \left(P_{5,6}^{(\mathrm{eq})}\Big|_{\mathcal{T}} + P_{7,8}^{(\mathrm{eq})}\Big|_{\mathcal{T}}\right)\frac{1}{(1+\exp(-\beta_e'))}. \quad (3)$$

In the above formulas, $P_{2j-1,2j}^{(\mathrm{eq})}\Big|_{\mathcal{T}} = \mathcal{P}\exp\left(-E_j/k_B\mathcal{T}\right)$ denotes the Boltzmann population at temperature $\mathcal{T}$, $E_j$ is the electronic energy in each pair of states $j = 1 \cdots 6$, and $\mathcal{P}$ is a normalization constant. Eqs. (2) and (3) express the fact that fractional populations within the $|m_S = 0\rangle$ and $|m_S = -1\rangle$ manifolds reorganize independently after the jump to attain a transient spin temperature, different for each manifold[15]. Therefore, assuming the initial (equilibrium) nuclear polarization is negligible, an rf-induced exchange of the populations in states $|4\rangle$ and $|5\rangle$ yields

$$P_n \approx P_{1,2}^{(\mathrm{eq})}\Big|_{\mathcal{T}}\frac{(1+\exp(-\beta_e))}{(1+\exp(-\beta_e'))}\left(\exp(-\beta_e') - \exp(-\beta_e)\right), \quad (4)$$

where $\beta_e \equiv |\gamma_e|B/k_B\mathcal{T}$. In deriving these expressions, we note that a spin temperature description is warranted at all times during the thermal jump given the comparatively short correlation times of the phonon bath (here serving as the "lattice"[19]).

In the limit where $\exp(-\beta_e), \exp(-\beta_e') \ll 1$, Eq. (4) boils down to $P_n \approx \exp(-\beta_e') - \exp(-\beta_e)$. On the other hand, a thermal jump where $\exp(-\beta_e) \sim 0$ ($\exp(-\beta_e) \sim 1$) and $\exp(-\beta_e') \sim 1$ ($\exp(-\beta_e') \sim 0$) yields the limit warm-up (cool-down) nuclear polarization $\left|P_{n\uparrow}^{(\max)}\right| = 1/2$ ($\left|P_{n\downarrow}^{(\max)}\right| = 1/3$). The asymmetry — still present in intermediate cases, see Fig. 2d — stems from the initial population trapped in the $|m_S = +1\rangle$ manifold, nearly null in one case, or approaching $1/3$ of the total in the other. Nuclear polarization of the same magnitude but reversed sign results if the frequencies of the rf pulses in Fig. 2d are chosen in the opposite order. Further, while the above discussion assumes $T_{1e}^{(S')} \ll T_{1e}^{(S)}$, identical results follow in the opposite case provided the rf-pulse frequency changes to exchange the populations of states $|3\rangle$ and $|6\rangle$.

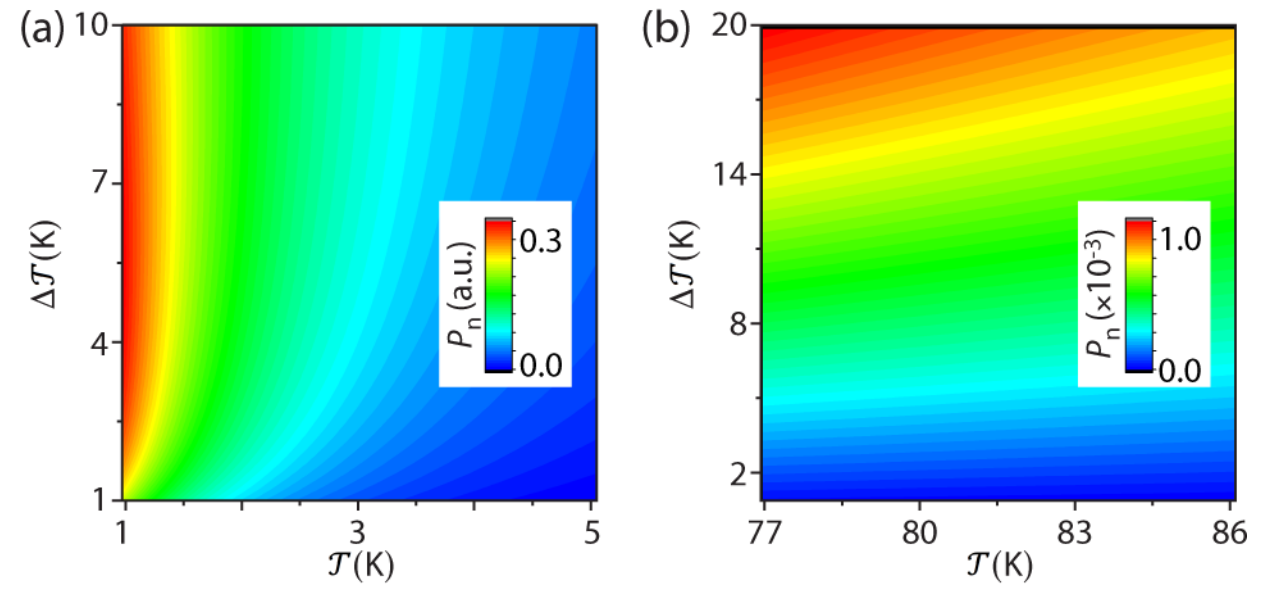

**FIG. 3: Thermal dependence.** (a) Nuclear spin polarization as a function of the base temperature $\mathcal{T}$ and thermal jump amplitude $\Delta\mathcal{T} = \mathcal{T}' - \mathcal{T}$ for the spin set in Fig. 2. In these calculations, we set $T_{1\mathrm{e}}^{(S)} = 5T_{1\mathrm{e}}^{(S')} = 50$ ms, $|\gamma_{\mathrm{e}}|B = 50$ GHz. (b) Same as above but near 80K, corresponding to liquid nitrogen temperatures.

Although the above considerations apply exclusively to the hyperfine coupled spin, repeated application of the protocol accompanied by spin diffusion to bulk nuclei — via inter-nuclear couplings or mediated via electron spin interactions[20,21] — will subsequently lead to a net accumulation of nuclear magnetization throughout the sample. The end level of polarization emerges from an interplay between the thermal cycle frequency, the polarization efficiency per cycle, and the spin-lattice relaxation time of bulk nuclei[22]. Note that because Eq. (4) derives exclusively from changes in $\beta_{\mathrm{e}}$, analogous dynamics can be attained if the lattice temperature remains constant and the applied magnetic field cycles between two magnetic fields $B$, $B'$ provided one of the two is proximal to $B_{\mathrm{m}}$. We emphasize that either version of the protocol ultimately relies on the difference between the spin-lattice relaxation times of spins $S$ and $S'$. The implication is that a thermal (or field) jump is fruitless in a system where the polarization source is a single electron spin coupled to a neighboring nucleus, the classical model in DNP

To better appreciate the practical implications of our approach, it is convenient to draw a comparison with existing nuclear polarization methods. Current DNP technology optimizes polarization gain through a protocol where the sample — prepared to contain a polarizing agent such as TEMPO or related nitroxides[23,24] — is initially frozen to cryogenic temperatures, irradiated with microwave at a moderately strong magnetic field, and then shuttled to a second superconducting magnet for inspection (often after flash-thawing). To mitigate the need for sample shuttling, recent efforts have been directed to developing methods adapted to stronger magnetic fields[25]. Work in this front articulates the synthesis of suitable polarizing agents[26,27] and the development of DNP sequences tailored to bring down the required microwave power to a minimum[28].

Without the need for high-power microwave generation — a demanding task at high magnetic fields typically requiring a gyrotron — our approach can nonetheless benefit from the use of cryogenic conditions, because large population changes — and hence high nuclear polarizations — result from even small temperature jumps. This is shown in Figs. 3a and 3b where we plot the end nuclear polarization in the spin set of Fig. 2 as a function of the thermal jump amplitude $\Delta\mathcal{T} = \mathcal{T}' - \mathcal{T}$ for different base temperatures $\mathcal{T}$. Note that $\Delta\mathcal{T} = \mathcal{T} = 1.5$ K corresponds to conditions entirely within the operating temperature range of present dissolution DNP experiments[29]. On the other hand, $P_{\mathrm{n}} \approx 10^{-3}$ with a 20 K jump near 80 K, approximately 25 (100) times the $^{1}$H ($^{13}$C) polarization at 1.8 T at these temperatures.

Under the cryogenic conditions assumed above, the spin-lattice relaxation times of typical paramagnetic moieties reach (and often exceed[30,31]) 100 ms, implying that the time interval for a thermal jump can be substantial. For example, bath heating could be quickly enacted by sample illumination with an infrared laser. In particular, we find large end nuclear spin polarization even if $T_{1\mathrm{e}}^{(S)}$, $T_{1\mathrm{e}}^{(S')}$ differ by as little as a factor 2, and the finite duration of the jump $\Delta t_{\updownarrow}$ is comparable to $T_{1\mathrm{e}}^{(S')}$.[15] As a reference, both heating and cooling jumps with a 50 K amplitude have been recently attained on a 1 µs time scale in the context of protein folding studies[32]. While this protocol is primarily conceived for nuclear magnetic resonance studies of organic materials, initial demonstrations could benefit from select inorganic platforms. One example is diamond, an excellent thermal conductor where co-existing point defects such as the NV and P1 centers — respectively featuring spin numbers $S = 1$ and $S' = 1/2$ — are known to exhibit different spin-lattice relaxation times[33].

Since the energy level structure near energy matching is largely independent of $B$, we predict only moderate changes in the polarization efficiency at high magnetic fields provided the crystal field also grows proportionally to shift the anti-crossing. Further, because the hyperfine couplings of nuclei within the first few atomic shells around a paramagnetic center are large (e.g., 10–150 MHz), practical effective Rabi amplitudes can be attained throughout the range of magnetic fields typical in nuclear magnetic resonance. In particular, it can be shown that, for a fixed thermal jump amplitude, the optimum matching field can be shifted to higher values by raising, not lowering, the base temperature[15].

In summary, we introduced a microwave-free route to dynamic nuclear polarization that builds on the transient imbalance between electron spin populations in nearly-degenerate spin levels upon a rapid thermal jump. The end nuclear polarization grows with the difference between the electron spin-lattice relaxation times of the two paramagnetic centers present in the polarizing agent to reach a limit value equal to $1/2$ if the thermal jump amplitude is sufficiently large (though the polarization efficiency drops in the presence of imperfect rf-pulses or heterogeneous broadening[15]). Optimal gains can be attained under cryogenic conditions through current DNP instrumentation adapted to produce thermal cycles of small (~1-10 K) amplitude. Operation at high magnetic fields is possible if one of the paramagnetic centers in the polarizing agent is designed to feature large crystal fields. This could be attained, for example, through the use of molecules pairing a radical and a rare-earth ion, where zero field splittings of up to ~10 THz are possible[34]. While this work focused on the response under cryogenic temperatures, thermal cycling near ambient conditions may also prove worth exploring. For example, the calculated nuclear polarization after a 50 K jump above room temperature in a 10 T field amounts to $P_{\mathrm{n}} = \frac{|\gamma_e| B}{6 k_{\mathrm{B}} \mathcal{T}} \frac{\Delta \mathcal{T}}{\mathcal{T}} \approx 1.5\times 10^{-3}$, approximately a factor 30 (120) above the equilibrium $^{1}$H ($^{13}$C) polarization at this field[15].

Work by C.A.M. was supported by the U.S. Department of Energy (DOE), Office of Science, Basic Energy Sciences (BES) under Award BES-DE-SC0020638; he also acknowledges access to the facilities and research infrastructure of the NSF CREST IDEALS, grant number NSF-HRD-1547830. P.R.Z. acknowledges financial support from SeCyT-UNC (33620180100154CB).

# Supplementary Material for:

# Microwave-free dynamic nuclear polarization via sudden thermal jumps

Carlos A. Meriles[1,2,†] and Pablo R. Zangara[3,4]

[1]*Department. of Physics, CUNY-City College of New York, New York, NY 10031, USA.*
[2]*CUNY-Graduate Center, New York, NY 10016, USA.*
[3]*Universidad Nacional de Córdoba, Facultad de Matemática, Astronomía, Física y Computación, Córdoba, Argentina.*
[4] *CONICET, Instituto de Física Enrique Gaviola (IFEG), Córdoba, Argentina.*

[†]*Corresponding author. E-mail: cmeriles@ccny.cuny.edu.*

## I. The spin Hamiltonian

The spin Hamiltonian describing the three-spin system we consider is

$$H = \Delta S_z^2 - \gamma_{\mathrm{e}} \mathbf{B} \cdot \mathbf{S} - \gamma_{\mathrm{e}} \mathbf{B} \cdot \mathbf{S}' - \gamma_{\mathrm{n}} \mathbf{B} \cdot \mathbf{I} + H_{\mathrm{d}}(\mathbf{S}, \mathbf{S}') + \mathbf{S} \cdot \mathbf{A} \cdot \mathbf{I} \,. \qquad (\mathrm{A.1})$$

Here, the first term describes the zero-field splitting $\Delta$ of the $S = 1$ spin, and we use it to define the z-direction of our reference frame. The second, third, and fourth terms respectively correspond to the Zeeman energies of the electronic spin $S$, the electronic spin $S' = 1/2$, and the nuclear spin $I = 1/2$. In what follows we assume that the external magnetic field is aligned with the z-direction as determined by the crystal field.

The dipolar interaction between spins $S$ and $S'$ is represented by $H_{\mathrm{d}}(\mathbf{S}, \mathbf{S}')$ and it can be written as:

$$H_{\mathrm{d}}(\mathbf{S}, \mathbf{S}') = H_{\mathrm{d}}^{(a)} + H_{\mathrm{d}}^{(b)} + H_{\mathrm{d}}^{(c)} + H_{\mathrm{d}}^{(d)} \,, \qquad (\mathrm{A.2})$$

where

$$H_{\mathrm{d}}^{(a)} = [1 - 3(\cos\theta)^2] S_z S_z'$$

$$H_{\mathrm{d}}^{(b)} = -\frac{1}{4}[1 - 3(\cos\theta)^2](S_+ S_-' + S_- S_+')$$

$$H_{\mathrm{d}}^{(c)} = -\frac{3}{2}\cos\theta \sin\theta \left[e^{-i\phi}(S_z S_+' + S_+ S_z') + e^{+i\phi}(S_z S_-' + S_- S_z')\right]$$

$$H_{\mathrm{d}}^{(d)} = -\frac{3}{4}(\sin\theta)^2 \left[e^{-2i\phi} S_+ S_+' + e^{+2i\phi} S_- S_-'\right]$$

Here, $(\theta, \phi)$ are the spherical coordinates of the inter-spin vector relative to the reference frame defined above. At any arbitrary magnetic field, the natural asymmetry of the two interacting spins (being $S = 1$ and $S' = 1/2$) leads to a hard energy mismatch that ultimately truncates all off-diagonal transition matrix elements induced by $H_{\mathrm{d}}^{(b)}$, $H_{\mathrm{d}}^{(c)}$ or $H_{\mathrm{d}}^{(d)}$. This implies that the only relevant term in $H_{\mathrm{d}}(\mathbf{S}, \mathbf{S}')$ is $H_{\mathrm{d}}^{(a)} \propto S_z S_z'$. Nevertheless, in the special case where $2|\gamma_{\mathrm{e}}|B \approx \Delta$, the double-quantum terms $H_{\mathrm{d}}^{(d)}$ become effectively *secular* or resonant since the electronic spin states $|m_S = 0, m_{S'} = 1/2\rangle$ and $|m_S = -1, m_{S'} = -1/2\rangle$ are nearly degenerate. In our simulations, we consider a *vicinity* of this degeneracy condition, so we retain $H_{\mathrm{d}}^{(d)}$ as it induces a (weak-) hybridization of spin states. It is important to stress that, in our simulations, we avoid working at the precise degeneracy point.

The last term in Eq. (A.1) refers to the hyperfine coupling between the electronic spin $S$ and the nuclear spin $I$. In general, for an arbitrary hyperfine tensor $\mathbf{A}$, the large difference between electronic and nuclear energy scales ($|\gamma_\mathrm{e}| \gg \gamma_\mathrm{n}$) leads to the secularization of this interaction, where those terms involving electronic spin-flips are disregarded,

$$\mathbf{S} \cdot \mathbf{A} \cdot \mathbf{I} = \sum_{\mu,\nu \in \{x,y.z\}} A_{\mu\nu} S_\mu I_\nu \approx A_{zz} S_z I_z + A_{zx} S_z I_x \,. \tag{A.3}$$

Under all these conditions, Eq. (A.1) boils down to

$$H = \Delta S_z^2 - \gamma_\mathrm{e} B S_z - \gamma_\mathrm{e} B S_z' - \gamma_\mathrm{n} B I_z + \mathcal{J}_\mathrm{d}[S_z S_z' + \zeta(S_+ S_+' + S_- S_-')] + A_{zz} S_z I_z + A_{zx} S_z I_x \,, \tag{A.4}$$

where $\mathcal{J}_\mathrm{d}$ sets the global scale of the dipolar coupling, and $\zeta \sim \mathcal{O}(1)$ is the appropriate angular pre-factor.

## II. Temperature jumps and relaxation-driven dynamics

We derive here a simplified model for spin relaxation following a temperature jump. We assume that the spin-lattice relaxation process is independent for each of the spins involved and, in particular, $T_{1\mathrm{e}}^{(S')} \ll T_{1\mathrm{e}}^{(S)}$. This implies that fractional populations within the $|m_S = 0\rangle$ and $|m_S = -1\rangle$ manifolds reorganize independently after the jump. Furthermore, we simplify our analysis by assuming no explicit temperature dependence on any of the $T_1$ times.

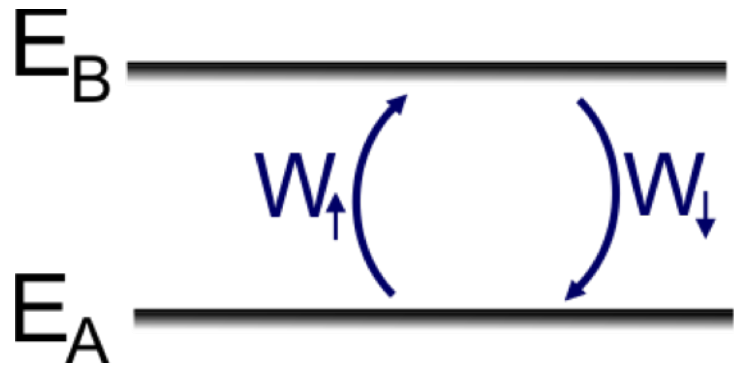

**Figure S1**. A TLS with energies $E_A < E_B$ and transition rates $W_\downarrow, W_\uparrow$.

We start by considering a generic two-level system (TLS), with two states $A$ and $B$, differing in energy as $\Delta E = E_B - E_A > 0$. As depicted in Fig. S1, such a TLS interacts with an external reservoir characterized by some temperature, and therefore provides a starting model for thermalization.

The TLS is assumed to be initially populated with $N = P_A + P_B$. Interaction with the reservoir at temperature $\mathcal{T}'$ ensures a population difference given by

$$P_A - P_B = N \left( \frac{W_\downarrow - W_\uparrow}{W_\downarrow + W_\uparrow} \right) . \tag{A.5}$$

Here, the transition rates are defined as

$$W_\downarrow = \frac{1}{T_1 \left[ 1 + exp\left\{ -\frac{\hbar \Delta E}{k_B \mathcal{T}'} \right\} \right]} \,. \tag{A.6}$$

$$W_\uparrow = \frac{1}{T_1 \left[ 1 + exp\left\{ +\frac{\hbar \Delta E}{k_B \mathcal{T}'} \right\} \right]} \,, \tag{A.7}$$

where $T_1 = (W_\downarrow + W_\uparrow)^{-1}$ defines the relaxation timescale. As in the main text, we simplify the notation through the definition $\beta' = \hbar \Delta E / (k_B \mathcal{T}')$. Then,

$$P_A = N\left(1 + e^{-\beta'}\right)^{-1} , \tag{A.8}$$

$$P_B = N\left(1 + e^{+\beta'}\right)^{-1} . \tag{A.9}$$

Now we recall the 12-level system shown in Fig. 2 of the main text. First, we consider a positive temperature jump, where the system is driven from equilibrium at temperature $\mathcal{T}$ to temperature $\mathcal{T}'$, where $\mathcal{T} < \mathcal{T}'$. This driving occurs within a timescale given by $T_{1\mathrm{e}}^{(S')}$.

Since $T_{1\mathrm{e}}^{(S')} \ll T_{1\mathrm{e}}^{(S)}$, the spin-lattice relaxation of $S'$ within the subset of states {1,2,3,4} occurs while keeping spin $S$ in the quantum state $|m_S = 0\rangle$. Since nuclear relaxation can be neglected within the considered timescale, we can identify the pair of states {1,2} as the single state $A$ introduced above and $\{3,4\}$ as the single state $B$. Following a time $t_\gamma^{(-)} \approx n_\gamma \times T_{1\text{e}}^{(S')}$, $n_\gamma \gtrsim 1$, after the temperature jump, the integrated transient populations $P_{1,2}^{(\text{tr})}$ and $P_{3,4}^{(\text{tr})}$ follow from Eqs. (A.8-A.9)

$$P_{1,2}^{(\text{tr})} = \left(P_{1,2}^{(\text{eq})}\Big|_{\mathcal{T}} + P_{3,4}^{(\text{eq})}\Big|_{\mathcal{T}}\right)\left(1 + e^{-\beta_e'}\right)^{-1}, \tag{A.10}$$

$$P_{3,4}^{(\text{tr})} = \left(P_{1,2}^{(\text{eq})}\Big|_{\mathcal{T}} + P_{3,4}^{(\text{eq})}\Big|_{\mathcal{T}}\right)\left(1 + e^{+\beta_e'}\right)^{-1}. \tag{A.11}$$

Here, $\beta_\text{e}' \equiv |\gamma_e| B / k_\text{B}\mathcal{T}'$. Note that the pre-factor $\left(P_{1,2}^{(\text{eq})}\Big|_{\mathcal{T}} + P_{3,4}^{(\text{eq})}\Big|_{\mathcal{T}}\right)$ plays the role of the initial population $N$ in the subset of states.

The same argument can be used in the subset of states $\{5,6,7,8\}$ characterized by $|m_S = -1\rangle$, with the association of states $\{5,6\}$ as state $A$ and states $\{7,8\}$ as $B$. Thus,

$$P_{5,6}^{(\text{tr})} = \left(P_{5,6}^{(\text{eq})}\Big|_{\mathcal{T}} + P_{7,8}^{(\text{eq})}\Big|_{\mathcal{T}}\right)\left(1 + e^{-\beta_e'}\right)^{-1}, \tag{A.12}$$

$$P_{7,8}^{(\text{tr})} = \left(P_{5,6}^{(\text{eq})}\Big|_{\mathcal{T}} + P_{7,8}^{(\text{eq})}\Big|_{\mathcal{T}}\right)\left(1 + e^{+\beta_e'}\right)^{-1}. \tag{A.13}$$

Defining the nuclear polarization as $P_\text{n} = \sum_{j=1}^{6}\left(P_{2j-1} - P_{2j}\right)$ (note the reversal in the state labels of Fig. 2a within the $|m_S = +1\rangle$ manifold) and assuming a negligible initial value, the exchange of populations between states $|4\rangle$ and $|5\rangle$ creates an imbalance which yields a net polarization. In fact, before the swapping operation at $t_\gamma^{(-)}$,

$$P_\text{n} = P_3^{(\text{tr})} - P_4^{(\text{tr})} + P_5^{(\text{tr})} - P_6^{(\text{tr})} \approx 0\,, \tag{A.14}$$

while after the swapping operation at $t_\gamma^{(+)}$ the roles of $P_4^{(\text{tr})}$ and $P_5^{(\text{tr})}$ are exchanged,

$$P_\text{n} = P_3^{(\text{tr})} + P_4^{(\text{tr})} - P_5^{(\text{tr})} - P_6^{(\text{tr})} = P_{3,4}^{(\text{tr})} - P_{5,6}^{(\text{tr})}. \tag{A.15}$$

Then, using Eqs. (A.11) and (A.12) we get

$$P_\text{n} = P_{3,4}^{(\text{tr})} - P_{5,6}^{(\text{tr})} = \left(P_{1,2}^{(\text{eq})}\Big|_{\mathcal{T}} + P_{3,4}^{(\text{eq})}\Big|_{\mathcal{T}}\right)\left(1 + e^{+\beta_e'}\right)^{-1} - \left(P_{5,6}^{(\text{eq})}\Big|_{\mathcal{T}} + P_{7,8}^{(\text{eq})}\Big|_{\mathcal{T}}\right)\left(1 + e^{-\beta_e'}\right)^{-1} \tag{A.16}$$

$$= \left[P_{1,2}^{(\text{eq})}\Big|_{\mathcal{T}} e^{-\beta_e'} + P_{3,4}^{(\text{eq})}\Big|_{\mathcal{T}} e^{-\beta_e'} - P_{5,6}^{(\text{eq})}\Big|_{\mathcal{T}} - P_{7,8}^{(\text{eq})}\Big|_{\mathcal{T}}\right]\left(1 + e^{-\beta_e'}\right)^{-1}. \tag{A.17}$$

In addition, the equilibrium condition following from Eqs. (A.8) and (A.9) implies that

$$\left(P_{3,4}^{(\text{eq})}\Big|_{\mathcal{T}} / P_{1,2}^{(\text{eq})}\Big|_{\mathcal{T}}\right) = \left(P_{5,6}^{(\text{eq})}\Big|_{\mathcal{T}} / P_{1,2}^{(\text{eq})}\Big|_{\mathcal{T}}\right) = \left(P_{7,8}^{(\text{eq})}\Big|_{\mathcal{T}} / P_{5,6}^{(\text{eq})}\Big|_{\mathcal{T}}\right) = e^{-\beta_e}\,. \tag{A.18}$$

Then,

$$P_\text{n} = P_{1,2}^{(\text{eq})}\Big|_{\mathcal{T}} \frac{\left(1 + e^{-\beta_e}\right)}{\left(1 + e^{-\beta_e'}\right)}\left[e^{-\beta_e'} - e^{-\beta_e}\right], \tag{A.19}$$

which corresponds to Eq. (4) in the main text. In the 'low temperature limit', $\exp(-\beta_\text{e}), \exp(-\beta_\text{e}') \ll 1$ and thus only the states $\{1,2\}$ are populated, $P_{1,2}^{eq}\big|_{\mathcal{T}} \approx 1$. This implies that

$$P_\text{n} \xrightarrow[\beta_\text{e},\beta_\text{e}' \to \infty]{} \exp(-\beta_\text{e}') - \exp(-\beta_\text{e})\,. \tag{A.20}$$

Notice that this last quantity is positive since $\beta_\text{e}' < \beta_\text{e}$. Furthermore, in the limit case where $\exp(-\beta_\text{e}) \ll 1$ but $\exp(-\beta_\text{e}') \approx 1$, the populations satisfy $P_{1,2}^{(\text{eq})}\Big|_{\mathcal{T}} \approx 1$ and thus

$$P_\text{n} \xrightarrow[\substack{\beta_\text{e} \to \infty \\ \beta_\text{e}' \to 0}]{} \frac{1}{2}\,. \tag{A.21}$$

Let us now consider a negative temperature jump, where the system is driven from temperature $\mathcal{T}$ to temperature $\mathcal{T}'$, with $\mathcal{T} > \mathcal{T}'$. In this case, the populations after a time at $t_\gamma^{(-)}$ following the jump are formally the same as stated in Eqs. (A.10-A.13), but the population exchange or swap here involves states $|3\rangle$ and $|6\rangle$. The equivalent of Eq. (A.15) is:

$$P_{\mathrm{n}} = -P_3^{(\mathrm{tr})} - P_4^{(\mathrm{tr})} + P_5^{(\mathrm{tr})} + P_6^{(\mathrm{tr})} = P_{5,6}^{(\mathrm{tr})} - P_{3,4}^{(\mathrm{tr})}\,, \tag{A.22}$$

since now populations $P_3^{(\mathrm{tr})}$ and $P_6^{(\mathrm{tr})}$ are exchanged. Then,

$$P_{\mathrm{n}} = P_{1,2}^{(\mathrm{eq})}\Big|_{\mathcal{T}} \frac{\left(1+e^{-\beta_e}\right)}{\left(1+e^{-\beta_e'}\right)}\left[-e^{-\beta_e'} + e^{-\beta_e}\right]. \tag{A.23}$$

In this case, the 'low temperature limit' $\exp(-\beta_{\mathrm{e}}), \exp(-\beta_{\mathrm{e}}') \ll 1$, yields

$$P_{\mathrm{n}} \xrightarrow[\beta_{\mathrm{e}},\beta_{\mathrm{e}}'\to\infty]{} \exp(-\beta_{\mathrm{e}}) - \exp(-\beta_{\mathrm{e}}')\,. \tag{A.24}$$

Notice that this value, as in Eq. (A.20), remains positive since here $\beta_{\mathrm{e}} < \beta_{\mathrm{e}}'$. In addition, in the limit case where $\exp(-\beta_{\mathrm{e}}') \ll 1$ but $\exp(-\beta_{\mathrm{e}}) \approx 1$, we have $P_{1,2}^{(\mathrm{eq})}\Big|_{\mathcal{T}} \approx 1/6$ and finally

$$P_{\mathrm{n}} \xrightarrow[\substack{\beta_{\mathrm{e}}'\to\infty \\ \beta_{\mathrm{e}}\to 0}]{} \frac{1}{3}\,, \tag{A.25}$$

Figure S2 shows the limiting cases determined by Eqs. (A.21) and (A.25). Ideal conditions are assumed: We consider a single temperature jump with $T_{1\mathrm{e}}^{(S')} = 10$ ms and $T_{1\mathrm{e}}^{(S)}, T_{1\mathrm{n}} \to \infty$. The next section introduces a more comprehensive analysis of the parameter space towards a realistic protocol.

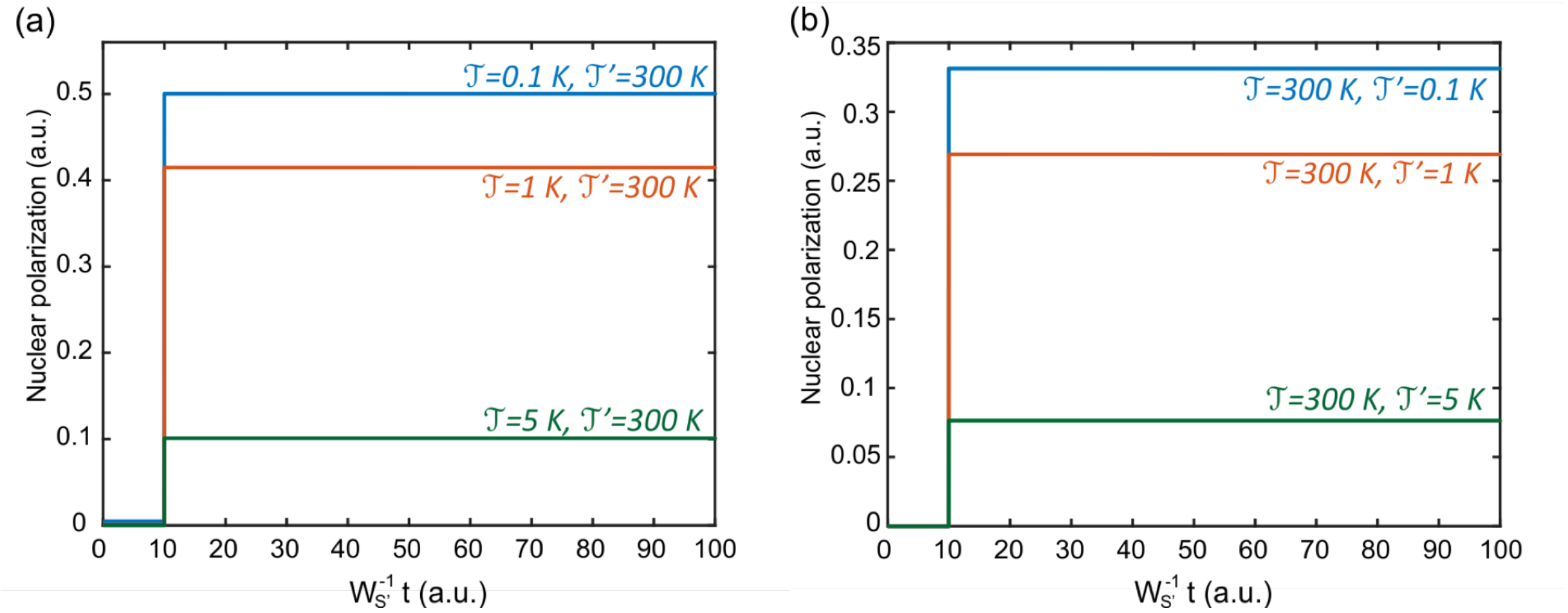

**Figure S2.** Limit cases for a single thermal jump under ideal conditions. (**a**) The system is heated from equilibrium at temperature $\mathcal{T}$ to temperature $\mathcal{T}'$ ($\mathcal{T} < \mathcal{T}'$). The upper limit of ~0.5 corresponds to Eq. (A.21). (**b**) The system is cooled from temperature $\mathcal{T}$ to temperature $\mathcal{T}'$ ($\mathcal{T} > \mathcal{T}'$). The upper limit of ~0.33 corresponds to Eq. (A.25). In both cases, $\Delta = 100$ GHz ($B \approx 1.8$ T), $W_{S'}^{-1} = T_{1\mathrm{e}}^{(S')} = 10$ ms, $T_{1\mathrm{e}}^{(S)} \equiv W_S^{-1} \to \infty$, $T_{1\mathrm{n}} \equiv W_I^{-1} \to \infty$, $\Delta t_{\uparrow} = \Delta t_{\downarrow} = T_{1\mathrm{e}}^{(S')}$ and the corresponding population exchange is assumed instantaneous at $t_\gamma \approx 10 \times T_{1\mathrm{e}}^{(S')}$.

## III. Exploring the parameter space

Any actual experimental protocol will deviate from the idealized jump cases discussed in the previous section. Specifically, some of the hypothesis underlying the arguments introduced above may have to be relaxed. First, thermal nuclear polarization cannot be beforehand neglected. Low temperature and strong magnetic fields would contribute to the observed nuclear polarization. Second, and more importantly, even though the nuclear relaxation timescale $T_{1n}$ can safely be assumed to be much longer than any of the electronic timescales $T_{1e}^{(S)}$, $T_{1e}^{(S')}$, the difference between these last two may not be considerably large. This means that simultaneous electronic relaxation can indeed happen. Third, an imperfect population exchange given by non-ideal $\pi$-pulses would in principle decrease the efficiency of each thermal jump. This observation is particularly relevant if the duration of these pulses becomes comparable to any of the relaxation timescales, since in such a case the assumption of an *instantaneous* swap operation breaks down.

The polarization protocol ends up operating in a non-ideal regime, where an appropriate fine-tuning of the sequence is required. Figure **S3** shows a concatenation of cooling and heating processes as the one shown in Fig. 2b of the main text. In principle, given the two electronic timescales $T_{1e}^{(S)}$, $T_{1e}^{(S')}$, the 'free' parameters to determine are:

1. The cycle time $\tau_c$ that accounts for the duration of a single heating or cooling process. In general, we choose $\tau_c$ to be many times the slowest electronic relaxation timescale, i.e. $\tau_{\mathrm{c}} \approx n_{\mathrm{c}} \times T_{1e}^{(S)}$, $n_{\mathrm{c}} > 1$.
2. The heating and cooling timescales, $\Delta t_\uparrow$ and $\Delta t_\downarrow$ respectively. In all our simulations, we assume they are tied to the fastest relaxation timescale, i.e. $\Delta t_\uparrow = \Delta t_\downarrow = T_{1e}^{(S')}$. Furthermore, we consider an exponential dependence of the temperature driving,
$$\mathcal{T}_{\mathrm{b}}(t) = \mathcal{T}_{initial} e^{-t/\Delta t_{\uparrow(\downarrow)}} + \mathcal{T}_{target}\left[1 - e^{-t/\Delta t_{\uparrow(\downarrow)}}\right]$$
Here, $\mathcal{T}_{initial}$ is the bath temperature when the driving starts and $\mathcal{T}_{target}$ is the desired final temperature. Notice that neither $\mathcal{T}_{initial}$ may be exactly the reference temperature ($\mathcal{T}_{hot}$ or $\mathcal{T}_{cold}$) nor $\mathcal{T}_{target}$ may be the actual 'final' temperature (the starting point for the next cycle). This is the case when $\tau_c$ is not sufficiently large as compared to $\Delta t_\uparrow$ and $\Delta t_\downarrow$.
3. The 'gate' time $t_\gamma$ elapsed from the starting point of the temperature driving to the trigger of the corresponding $\pi$-pulse (population exchange). As stated before, we choose $t_\gamma$ to be comparable to or greater than the fastest electronic relaxation timescale, i.e. $t_\gamma \approx n_\gamma \times T_{1e}^{(S')}$, $n_\gamma \gtrsim 1$.

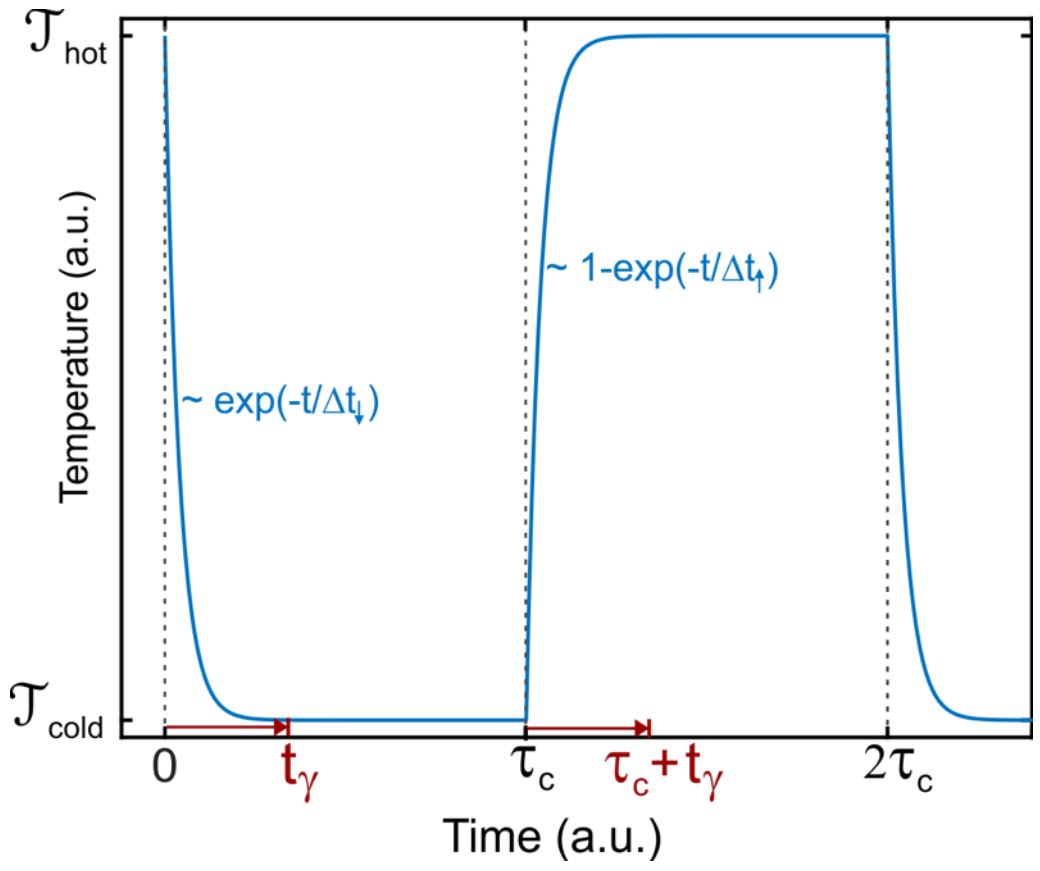

**Figure S3**. Relevant parameters in a basic sequence of concatenated cooling and heating cycles operating between two arbitrary reference temperatures, $\mathcal{T}_{hot}$ and $\mathcal{T}_{cold}$.

To simulate a protocol as introduced above, we evaluate the population dynamics of the 12-level system using a classical rate equation

$$\frac{\mathrm{d}}{\mathrm{dt}}\mathbf{P} = \mathbb{W}\,\mathbf{P}\,, \tag{A.26}$$

where $\mathbb{W}$ accounts for all possible relaxation processes. The temperature driving requires a discretization of the time domain, so evolution of the system is given by

$$\mathbf{P}(\mathrm{t}+\mathrm{dt}) = \exp\{\,\mathbb{W}(\mathrm{t})\,\mathrm{dt}\,\}\,\mathbf{P}(\mathrm{t}). \tag{A.27}$$

The time evolution is periodically interrupted by an instantaneous $\pi$-pulse that produces the required population exchange.

Figure **S4** shows the obtained nuclear polarization after many cycles as a function of the 'gate' time $t_\gamma$ and the relative difference between the electronic timescales $T_{1e}^{(S)}$, $T_{1e}^{(S')}$. Here, we force $t_\gamma < \tau_c = n_c T_{1e}^{(S)}$ ($n_c = 5$ for the present case) which stands for the cycle time limit. Notice that the asymmetry between the opposite thermal jumps can decrease the accumulated nuclear polarization if $t_\gamma$ is not optimal. This is the case when the electronic timescales are too close, so simultaneous electronic spin relaxation does take place (Fig. S4(b)). It can also occur if $t_\gamma$ is too long and becomes comparable to the slowest electronic relaxation timescale $T_{1e}^{(S)}$ (Fig. S4(c)).

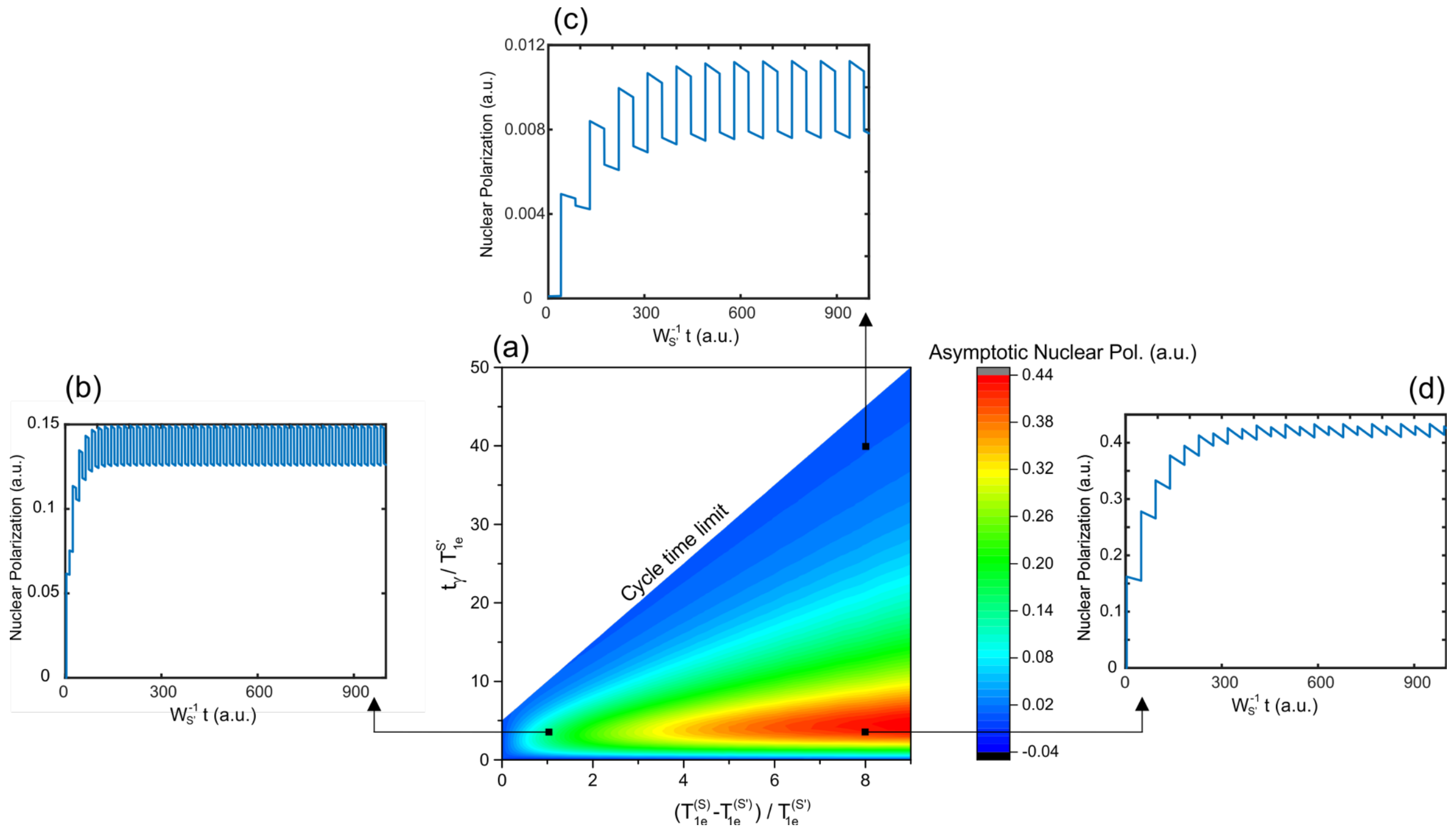

**Figure S4.** (**a**) Nuclear spin polarization as a function of the relative trigger time $t_\gamma/T_{1e}^{(S')}$ and fractional electronic spin-lattice relaxation time $\Delta T/T_{1e}^{(S')} \equiv \left(T_{1e}^{(S)} - T_{1e}^{(S')}\right)/T_{1e}^{(S')}$; here, we assume $\mathcal{T}_{hot} = 5$ K, $\mathcal{T}_{cold} = 1$ K, $\Delta t_\uparrow = \Delta t_\downarrow = T_{1e}^{(S')}$, $\tau_c = 5T_{1e}^{(S)}$, $W_{S'}^{-1} = T_{1e}^{(S')} = 10$ ms, $T_{1n} = 10$ s, $\Delta = 100$ GHz ($B \approx 1.8$ T). Explicit time-traces are shown in (**b**), (**c**) and (**d**).

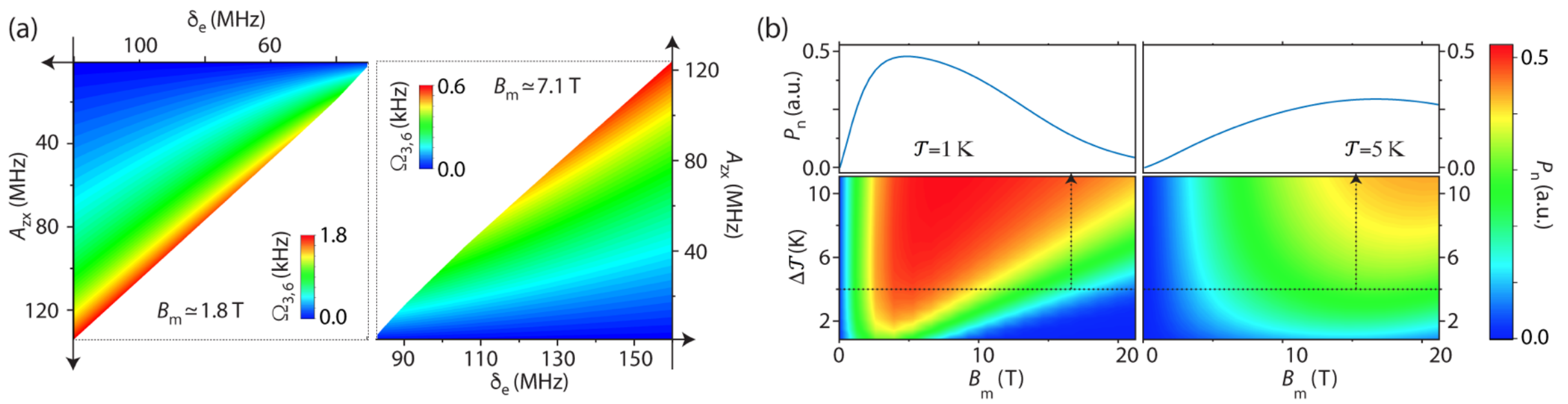

**FIG. S5: Dependence on applied magnetic field and inter-spin couplings.** (a) Effective interstate Rabi field $\Omega_{3,6}$ as a function of the field detuning $\delta_{\rm e}$ relative to matching, and the pseudosecular hyperfine coupling amplitude $A_{zx}$ at two different matching fields. Throughout these calculations $\Omega = 50$ kHz, $\delta_{\rm n} = 10$ MHz, and $\mathcal{J}_{\rm d} = 7$ MHz. (b) Nuclear spin polarization $P_{\rm n}$ as a function of $\Delta\mathcal{T}$ and $B_{\rm m}$ for base temperatures of 1 and 5 K (left and right plots, respectively). The upper inserts are cross sections at $\Delta\mathcal{T} = 4$ K. All other conditions as in Fig. 2 in the main text.

Now we turn our attention to the interplay between the efficiency of the rf pulses (or, equivalently, the required rf amplitude), the hyperfine coupling, and the operating magnetic field. Fig. S5a shows the effective Rabi amplitude $\Omega_{3,6}$ of the 'forbidden' transition between states $|3\rangle$ and $|6\rangle$ assuming a fixed rf amplitude $\Omega$. We find lower Rabi amplitudes for more-weakly coupled nuclei, consistent with the gradual truncation of the pseudosecular hyperfine coupling term $A_{zx}S_zI_x$ as the nuclear Zeeman contribution becomes dominant. On the other hand, because the hyperfine couplings of nuclei within the first few atomic shells around a paramagnetic center can be large (e.g., 10–150 MHz), practical effective Rabi amplitudes can be attained throughout the range of magnetic fields typical in nuclear magnetic resonance (see left and right panels in Fig. S5a). When assessing this response, it is worth emphasizing that the use of rf pulses is not mandatory as the protocol could equally exploit small dc field pulses (e.g., few gauss) designed to transiently match the energy levels required for the generation of nuclear polarization. We postpone the discussion of this strategy to future work.

Based on the above considerations, Fig. S5b shows the calculated nuclear polarization for variable $B_{\rm m}$ and thermal jumps of a few degrees. For a fixed thermal jump amplitude, $P_{\rm n}$ shows a maximum at a matching field whose value depends on $\mathcal{T}$. Interestingly, shifting the optimum to higher magnetic fields is possible by raising, not lowering, the base temperature (compare left and right panels in Fig. S5b), even though a thermal jump of greater amplitude is needed to maintain the polarization efficiency.

## IV. Polarization efficiency

As in other DNP mechanisms, the polarization transfer efficiency — which, thus far, we have assumed ideal — can be negatively impacted by multiple factors. While an in-depth analysis exceeds the scope of this work, here we consider two, namely, (*i*) the duration of the rf-pulse, and (*ii*) the heterogeneity of the crystal field splitting $\Delta$. To illustrate (*i*), we first consider the case where the rf-pulse is resonant with the $|4\rangle \leftrightarrow |5\rangle$ transition and assume the effect of the pulse amounts to a rotation of angle $\theta$ in the Bloch sphere defined by the two states (with $\theta = \pi$ describing the ideal case of population exchange assumed in the text). To this end, we write the system's density matrix immediately prior to the application of the pulse as

$$\rho^{(tr^-)} = \begin{bmatrix} P_4^{(tr^-)} & 0 \\ 0 & P_5^{(tr^-)} \end{bmatrix} = \frac{1}{2}\left(P_4^{(tr^-)} + P_5^{(tr^-)}\right)\mathbb{I} + \frac{1}{2}\left(P_4^{(tr^-)} - P_5^{(tr^-)}\right)\sigma_z\,, \qquad \text{(A.28)}$$

where we limit our description to the manifold spanned by states $|4\rangle$ and $|5\rangle$, and use $\mathbb{I}$ ($\sigma_{z,x,y}$) to denote the identity matrix (Pauli operator) in this manifold. Using the standard rotating frame approximation to describe the evolution of the system under the action of rf excitation, $\Omega\sigma_y \cos\omega_{4,5}t$, the density matrix after the pulse takes the form

$$\rho^{(tr^+)} = \tfrac{1}{2}\left(P_4^{(tr^-)} + P_5^{(tr^-)}\right)\mathbb{I} + \tfrac{1}{2}\left(P_4^{(tr^-)} - P_5^{(tr^-)}\right)(\sigma_z \cos\theta + \sigma_x \sin\theta)\,. \tag{A.29}$$

From Eq. (A.29), we express the populations after the pulse as

$$P_4^{(tr^+)} = \frac{1}{2}\left(P_4^{(tr^-)} + P_5^{(tr^-)}\right) + \frac{\cos\theta}{2}\left(P_4^{(tr^-)} - P_5^{(tr^-)}\right), \tag{A.30}$$

$$P_5^{(tr^+)} = \frac{1}{2}\left(P_4^{(tr^-)} + P_5^{(tr^-)}\right) - \frac{\cos\theta}{2}\left(P_4^{(tr^-)} - P_5^{(tr^-)}\right). \tag{A.31}$$

Using Eqs. (A.11) and (A.12) as well as (A.18), we write

$$P_3^{(tr^+)} = P_3^{(tr^-)} = \frac{1}{2}P_{1,2}^{(\mathrm{eq})}\Big|_{\mathcal{T}}\frac{\left(1 + e^{-\beta_\mathrm{e}}\right)}{\left(1 + e^{-\beta_e'}\right)}e^{-\beta_e'}\,, \tag{A.32}$$

$$P_4^{(tr^+)} = \frac{1}{4}P_{1,2}^{(\mathrm{eq})}\Big|_{\mathcal{T}}\frac{\left(1 + e^{-\beta_\mathrm{e}}\right)}{\left(1 + e^{-\beta_e'}\right)}\left(\left(e^{-\beta_e'} + e^{-\beta_\mathrm{e}}\right) + \cos\theta\left(e^{-\beta_e'} - e^{-\beta_\mathrm{e}}\right)\right), \tag{A.33}$$

$$P_5^{(tr^+)} = \frac{1}{4}P_{1,2}^{(\mathrm{eq})}\Big|_{\mathcal{T}}\frac{\left(1 + e^{-\beta_\mathrm{e}}\right)}{\left(1 + e^{-\beta_e'}\right)}\left(\left(e^{-\beta_e'} + e^{-\beta_\mathrm{e}}\right) - \cos\theta\left(e^{-\beta_e'} - e^{-\beta_\mathrm{e}}\right)\right), \tag{A.34}$$

$$P_6^{(tr^+)} = P_6^{(tr^-)} = \frac{1}{2}P_{1,2}^{(\mathrm{eq})}\Big|_{\mathcal{T}}\frac{\left(1 + e^{-\beta_\mathrm{e}}\right)}{\left(1 + e^{-\beta_e'}\right)}e^{-\beta_\mathrm{e}}\,. \tag{A.35}$$

Therefore, the nuclear polarization after the rf-pulse is given by

$$\begin{aligned} P_\mathrm{n} &= P_3^{(tr^+)} - P_6^{(tr^+)} + P_5^{(tr^+)} - P_4^{(tr^+)} = \\ &= \left(\sin\frac{\theta}{2}\right)^2 P_{1,2}^{(\mathrm{eq})}\Big|_{\mathcal{T}}\frac{\left(1 + e^{-\beta_\mathrm{e}}\right)}{\left(1 + e^{-\beta_e'}\right)}\left(e^{-\beta_e'} - e^{-\beta_\mathrm{e}}\right), \\ &= \left(\sin\frac{\theta}{2}\right)^2 P_\mathrm{n}^{(opt)}\,, \end{aligned} \tag{A.36}$$

where $P_\mathrm{n}^{(opt)}$ indicates the optimum calculated in Eq. (A.19). Note that $\theta = \pi/2$ yields half the maximum, implying that, in the absence of a proper rf amplitude calibration, polarization can be induced simply via rf saturation.

The possibility of non-ideal $\pi$-pulses can be further explored by considering not only a single thermal jump under ideal conditions, but by simulating many consecutive temperature cycles under imperfect conditions. To this end, we keep the assumption of an instantaneous manipulation of populations but we use Eq. (A.29) to describe the effect of rf manipulation in our simulations with $\theta$ as a free parameter.

Figure S6 shows an example where we compare the performance over many cycles for $\theta = 0, \frac{\pi}{4}, \frac{\pi}{2}, \frac{3\pi}{4}, \pi$. We confirm that nuclear polarization accumulates after many cycles even if the single-jump efficiency is not optimal. Notice that, as stated above, $\theta = \pi/2$ yields half the optimal polarization (corresponding to $\theta = \pi$) for the first thermal jump. But consecutive cycles can further improve the performance of the protocol, reaching almost the same saturation values. This reinforces our previous observation indicating that an exact calibration of the rf amplitude is unnecessary.

To assess the impact of heterogeneity in the crystal field, we consider the case of rf excitation at a fixed frequency $\omega_\mathrm{rf}$. Cases (a) through (d) in Fig. S7 illustrate four different situations where $\omega_\mathrm{rf}$ excites either the $|4\rangle \leftrightarrow |5\rangle$ transition (Figs. S7a and S7c) or the $|3\rangle \leftrightarrow |6\rangle$ transition (Figs. S7b ad S7d). The crystal field $\Delta$ required to attain energy matching in each case can be derived by considering the Hamiltonian in Eq. (A.1). Using $\omega_\mathrm{e}$ ($\omega_\mathrm{n}$) to denote the electronic (nuclear) Zeeman frequency, we obtain

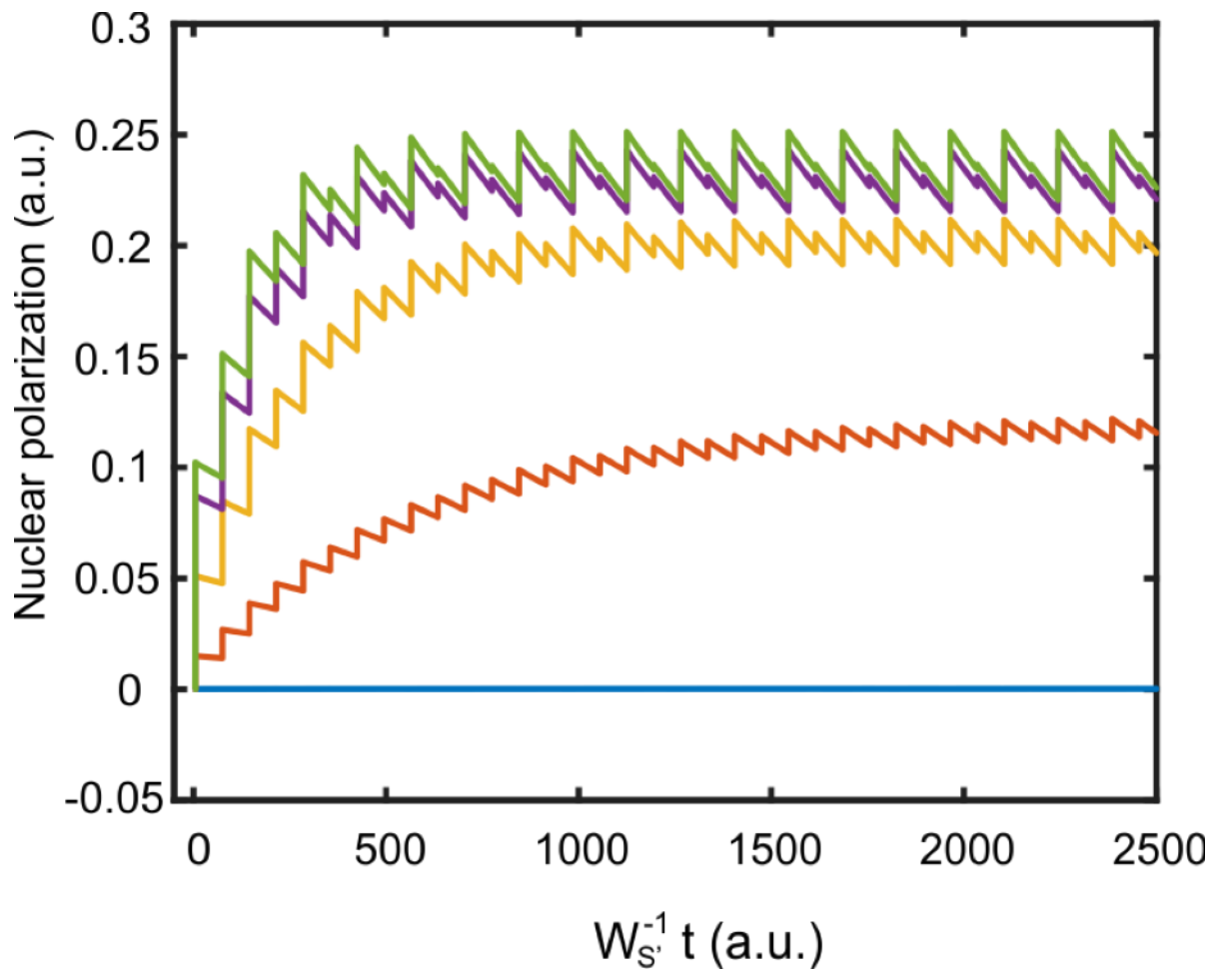

**Figure S6**. Nuclear polarization generated by non-ideal $\pi$-pulses as defined by Eq. (A.29), from top to bottom $\theta = \pi, \frac{3\pi}{4}, \frac{\pi}{2}, \frac{\pi}{4}, 0$. Here, $\mathcal{T}_{hot} = 10$ K, $\mathcal{T}_{cold} = 4$ K, $\Delta t_{\uparrow} = \Delta t_{\downarrow} = T_{1e}^{(S')}$, $W_S^{-1} = T_{1e}^{(S)} = 70$ ms, $W_{S'}^{-1} = T_{1e}^{(S')} = 10$ ms, $T_{1n} = 10$ s, $\tau_c = 10T_{1e}^{(S)}$, $t_\gamma = 5T_{1e}^{(S')}$, $\Delta = 300$ GHz ($B \approx 5.3$ T).

$$\Delta_{\mathrm{a}} = 2\omega_{\mathrm{e}} + \omega_{\mathrm{n}} + \omega_{\mathrm{rf}} + \frac{A_{zz}}{2}, \tag{A.37}$$

$$\Delta_{\mathrm{b}} = 2\omega_{\mathrm{e}} - \omega_{\mathrm{n}} + \omega_{\mathrm{rf}} - \frac{A_{zz}}{2}, \tag{A.38}$$

$$\Delta_{\mathrm{c}} = 2\omega_{\mathrm{e}} + \omega_{\mathrm{n}} - \omega_{\mathrm{rf}} + \frac{A_{zz}}{2}, \tag{A.39}$$

$$\Delta_{\mathrm{d}} = 2\omega_{\mathrm{e}} - \omega_{\mathrm{n}} - \omega_{\mathrm{rf}} - \frac{A_{zz}}{2}, \tag{A.40}$$

for cases (a) through (d), respectively. Therefore, in a heterogeneous distribution of crystal fields $f(\Delta)$, the resulting nuclear polarization is given by

$$P_{\mathrm{n}} = \delta f(\Delta, \omega_{\mathrm{rf}})\, P_{\mathrm{n}}^{(opt)}, \tag{A.41}$$

where $\delta f(\Delta, \omega_{\mathrm{rf}}) \equiv \big(f(\Delta_{\mathrm{a}}) + f(\Delta_{\mathrm{c}}) - f(\Delta_{\mathrm{b}}) - f(\Delta_{\mathrm{d}})\big)/\big(f(\Delta_{\mathrm{a}}) + f(\Delta_{\mathrm{c}}) + f(\Delta_{\mathrm{b}}) + f(\Delta_{\mathrm{d}})\big)$. Note that other similar scaling factors can be derived from considering other types of heterogeneity (e.g., in the case of a polycrystalline sample — where the crystal field axis takes random orientations — or when either electronic spin is affected by $g$-factor anisotropy). We also emphasize all these expressions are only valid in the limit where one electronic spin-lattice relaxation is much longer than the other.

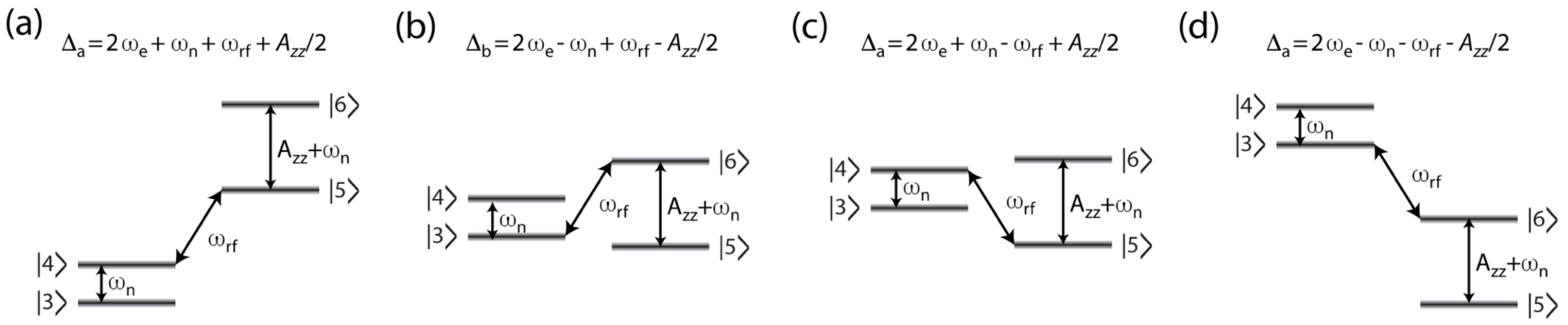

**Figure S7**. In the presence of crystal field heterogeneity, rf irradiation at a fixed frequency $\omega_{\mathrm{rf}}$ leads to different energy matching conditions represented in cases (a) through (d).

Therefore, from Eqs. (A.36) and (A.41), we conclude one can generically write the nuclear polarization as

$$P_{\mathrm{n}} = \varepsilon\eta\, P_{\mathrm{n}}^{(\mathrm{eq})}\Big|_{\mathcal{T}}\,, \tag{A.42}$$

where $\eta$ is an "interference" factor that takes into account cancellations between populations, $P_{\mathrm{n}}^{(\mathrm{eq})}\Big|_{\mathcal{T}}$ is the nuclear polarization in equilibrium at temperature $\mathcal{T}$, and the "enhancement" $\varepsilon$ is given by

$$\varepsilon = \frac{P_{1,2}^{(\mathrm{eq})}\Big|_{\mathcal{T}}}{P_{\mathrm{n}}^{(\mathrm{eq})}\Big|_{\mathcal{T}}} \frac{\left(1+e^{-\beta_{\mathrm{e}}}\right)}{\left(1+e^{-\beta_e'}\right)}\left(e^{-\beta_e'}-e^{-\beta_{\mathrm{e}}}\right). \tag{A.43}$$

## V. The high-temperature limit

Here we consider the case of thermal jumps in the limit of high temperatures where $\exp(-\beta_{\mathrm{e}}) \approx 1-\beta_{\mathrm{e}}$ and $\exp(-\beta_{\mathrm{e}}') \approx 1-\beta_{\mathrm{e}}'$. In this regime, we find

$$P_{1,2}^{(\mathrm{eq})} = \frac{1}{1+2e^{-\beta_{\mathrm{e}}}+e^{-2\beta_{\mathrm{e}}}+e^{-3\beta_{\mathrm{e}}}+e^{-4\beta_{\mathrm{e}}}} \approx \frac{1}{6}\left(1+11\beta_{\mathrm{e}}\right). \tag{A.44}$$

Similarly,

$$\frac{\left(1+e^{-\beta_{\mathrm{e}}}\right)}{\left(1+e^{-\beta_{\mathrm{e}}'}\right)} \approx 1-\frac{\left(\beta_{\mathrm{e}}+\beta_{\mathrm{e}}'\right)}{2}. \tag{A.45}$$

Therefore, the end nuclear spin polarization (Eq. (A.19)) takes the form

$$P_{\mathrm{n}} = P_{1,2}^{(\mathrm{eq})}\Big|_{\mathcal{T}} \frac{\left(1+e^{-\beta_e}\right)}{\left(1+e^{-\beta_e'}\right)}\left(e^{-\beta_e'}-e^{-\beta_e}\right) \approx \frac{1}{6}\left(1+11\beta_e\right)\left(1-\frac{\left(\beta_{\mathrm{e}}+\beta_{\mathrm{e}}'\right)}{2}\right)\left(\beta_{\mathrm{e}}-\beta_{\mathrm{e}}'\right), \tag{A.46}$$

or

$$P_{\mathrm{n}} \approx \frac{1}{6}\left(\beta_{\mathrm{e}}-\beta_{\mathrm{e}}'\right) \tag{A.47}$$

to first order in $\beta_{\mathrm{e}}$, $\beta_{\mathrm{e}}'$. Expressing the end temperature as $\mathcal{T}' = \mathcal{T}+\Delta\mathcal{T}$, we write $\beta_{\mathrm{e}}' = \frac{|\gamma_e|B}{k_{\mathrm{B}}(\mathcal{T}+\Delta\mathcal{T})} \approx \beta_{\mathrm{e}}\left(1-\frac{\Delta\mathcal{T}}{\mathcal{T}}\right)$ meaning that $\beta_{\mathrm{e}}-\beta_{\mathrm{e}}' \approx \beta_{\mathrm{e}}\frac{\Delta\mathcal{T}}{\mathcal{T}}$. Correspondingly,

$$P_{\mathrm{n}} \approx \frac{1}{6}\beta_{\mathrm{e}}\frac{\Delta\mathcal{T}}{\mathcal{T}} \tag{A.48}$$

as presented in the main text.